\documentclass[11pt, onecolumn]{IEEEtran} 
\pdfoutput=1
\usepackage{amsmath}
\usepackage{amssymb}
\usepackage{amsfonts}
\usepackage{epsfig}
\usepackage{subfigure}
\usepackage{calc}
\usepackage{color}
\usepackage[all]{xy}
\usepackage{bm}
\usepackage{enumerate}
\usepackage{hyperref}

\newtheorem{defn}{Definition}
\newtheorem{thm}{Theorem}[section]
\newtheorem{cor}[thm]{Corollary}
\newtheorem{prop}{Proposition}

\newtheorem{lem}[thm]{Lemma}
\newtheorem{conj}[thm]{Conjecture}
\newtheorem{constr}[thm]{Construction}
\newtheorem{note}{Remark}
\newtheorem{example}{Example}
\newcommand{\bit}{\begin{itemize}}
\newcommand{\eit}{\end{itemize}}
\newcommand{\bcor}{\begin{cor}}
\newcommand{\ecor}{\end{cor}}
\newcommand{\beq}{\begin{equation}}
\newcommand{\eeq}{\end{equation}}
\newcommand{\beqn}{\begin{equation*}}
\newcommand{\eeqn}{\end{equation*}}
\newcommand{\bea}{\begin{eqnarray}}
\newcommand{\eea}{\end{eqnarray}}
\newcommand{\bean}{\begin{eqnarray*}}
\newcommand{\eean}{\end{eqnarray*}}
\newcommand{\ben}{\begin{enumerate}}
\newcommand{\een}{\end{enumerate}}
\newcommand{\bdefn}{\begin{defn}}
\newcommand{\edefn}{\end{defn}}
\newcommand{\bnote}{\begin{note}}
\newcommand{\enote}{\end{note}}
\newcommand{\bprop}{\begin{prop}}
\newcommand{\eprop}{\end{prop}}
\newcommand{\blem}{\begin{lem}}
\newcommand{\elem}{\end{lem}}
\newcommand{\bthm}{\begin{thm}}
\newcommand{\ethm}{\end{thm}}
\newcommand{\bconj}{\begin{conj}}
\newcommand{\econj}{\end{conj}}
\newcommand{\bconstr}{\begin{constr}}
\newcommand{\econstr}{\end{constr}}
\newcommand{\bpf}{\begin{proof}}
\newcommand{\epf}{\end{proof}}

\begin{document}

\title{An Improved Outer Bound on the Storage-Repair-Bandwidth Tradeoff of Exact-Repair Regenerating Codes} 
 \author{Birenjith Sasidharan, Kaushik Senthoor and P. Vijay Kumar
 \\
Department of ECE, Indian Institute of Science, Bangalore, India.
\\
(email: \{biren,kaushik.sr,vijay\}@ece.iisc.ernet.in)
\thanks{This research is supported in part by the National Science Foundation under Grant 0964507 and in part by the NetApp Faculty Fellowship program.} }
\date{\today}
\maketitle

\begin{abstract}
In this paper we establish an improved outer bound on the storage-repair-bandwidth tradeoff of regenerating codes under exact repair.  The result shows that in particular, it is not possible to construct exact-repair regenerating codes that asymptotically achieve the tradeoff that holds for functional repair.   While this had been shown earlier by Tian for the special case of $[n,k,d]=[4,3,3]$ the present result holds for general $[n,k,d]$.  The new outer bound is obtained by building on the framework established earlier by Shah et al. 
\end{abstract}

\section{Introduction} \label{sec:introduction} 

In a distributed storage system, the data file comprising of $B$ data symbols drawn from a finite field $\mathbb{F}_q$, is encoded using  an error-correcting code of block length $n$ and the resulting code symbols are respectively stored in $n$ nodes of the storage network.   While repetition codes such as the triple replication commonly employed in a Hadoop Distributed File System (HDFS) \cite{hadoop} are extensively used, 
 there has been increasing interest lately in the storage industry for more sophisticated coding schemes that permit operation at low values of storage overhead.  Given the massive amount of data that is currently being stored, even a small reduction in storage overhead can translate into huge savings. For example, a large analytics cluster at Facebook could involve 3000 nodes that together store 230 million blocks, each of size 256 MB making for a total storage of several petabytes \cite{SatAstPapDimVadChe}.  
 
Quite apart from resiliency to node failure and reduced storage overhead there are several other attributes desirable in distributed storage system.  These include: 
 \bit
\item small repair bandwidth, i.e., the amount of data download in the case of a node failure is much smaller in comparison with the file size $B$, 
\item low repair degree, i.e., the number of helper nodes accessed for node repair is small, 
\item low update complexity, i.e., data can be updated in a low-complexity manner, 
\item the facilitating of data operations such as MapReduce.
\eit
An examples of a coding scheme currently used in practice is the $[14,10]$ Reed-Solomon (RS) code employed by Facebook in an open source module called HDFS RAID \cite{SatAstPapDimVadChe}.  However, RS codes have the disadvantage of requiring both large repair bandwidth as well as a large repair degree.  In response, the research community has come up with two recent alternatives to RS codes known respectively as regenerating codes \cite{DimGodWuWaiRam} and codes with locality \cite{GopHuaSimYek}.   The focus of the current paper is on regenerating codes. 

\section{Regenerating Codes} \label{sec:regenerating_codes} 

\begin{figure}[ht]
\begin{minipage}[b]{0.30\linewidth}
\centering
\includegraphics[height=1.5in]{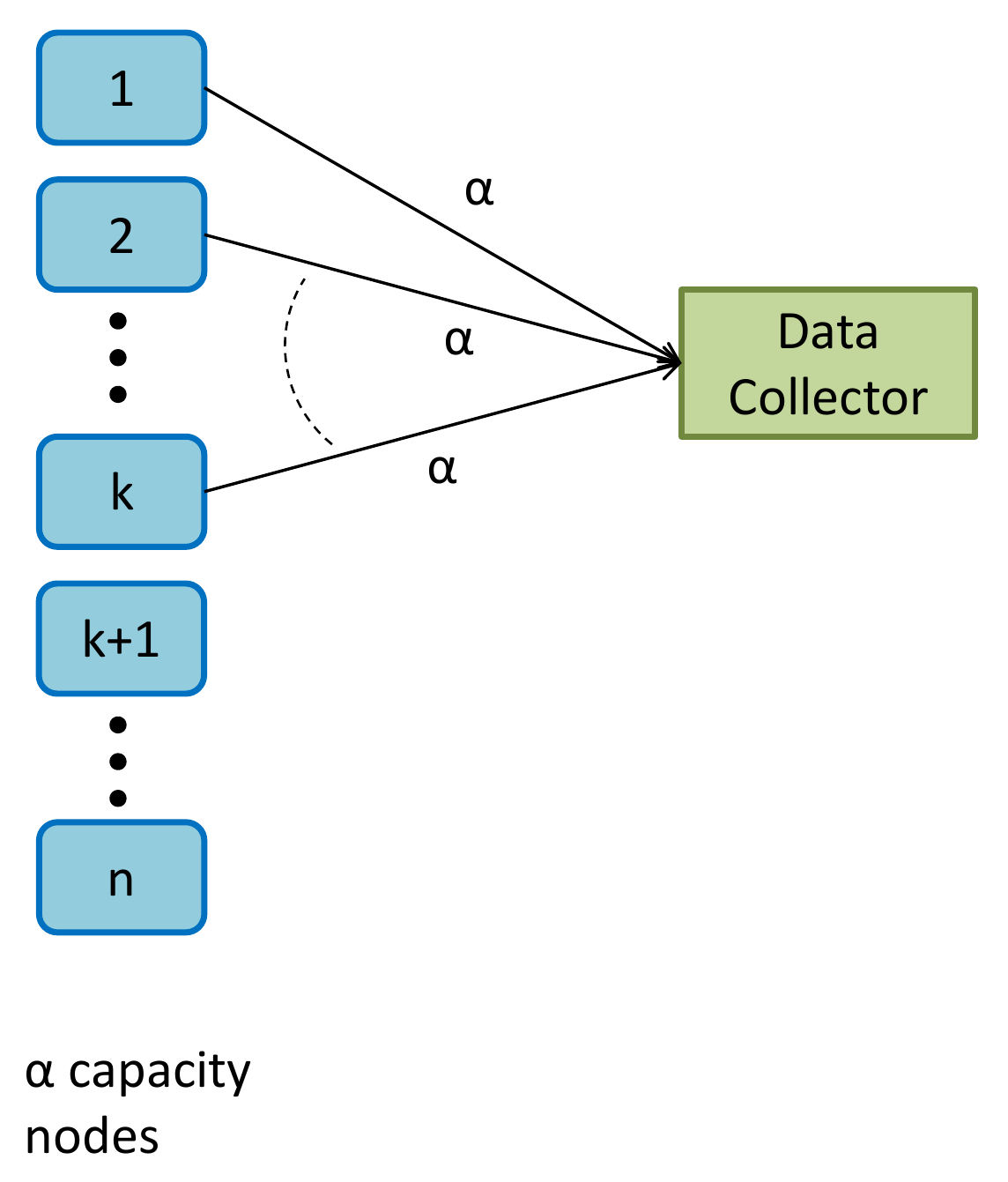}
\caption{Data collection.}
\label{fig:data_collection}
\end{minipage}
\hspace{-0.5cm}
\begin{minipage}[b]{0.30\linewidth}
\centering
\includegraphics[height=1.5in]{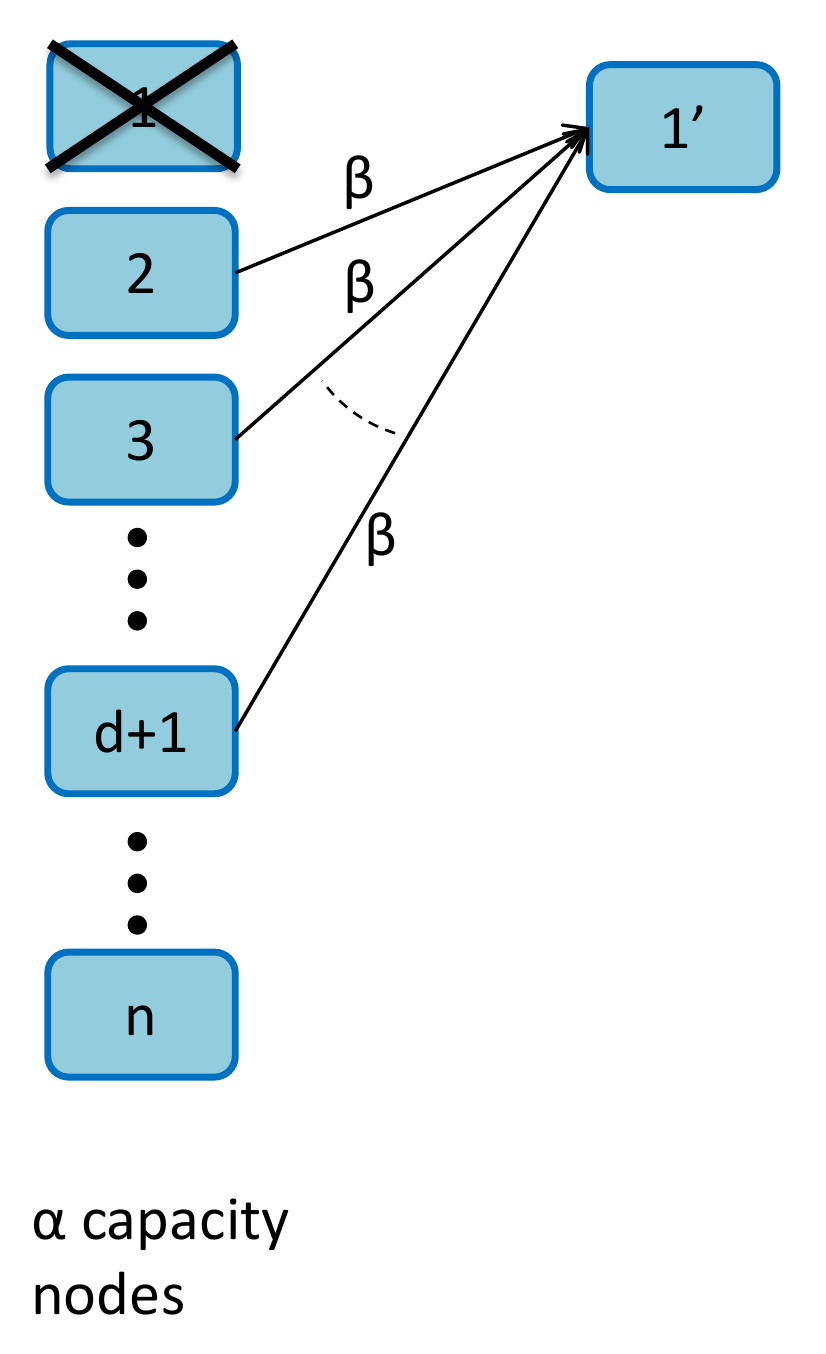}
\caption{Node repair.}
\label{fig:node_repair}
\end{minipage}
\hspace{-0.5cm}
\begin{minipage}[b]{0.30\linewidth}
\centering
\includegraphics[width=3in]{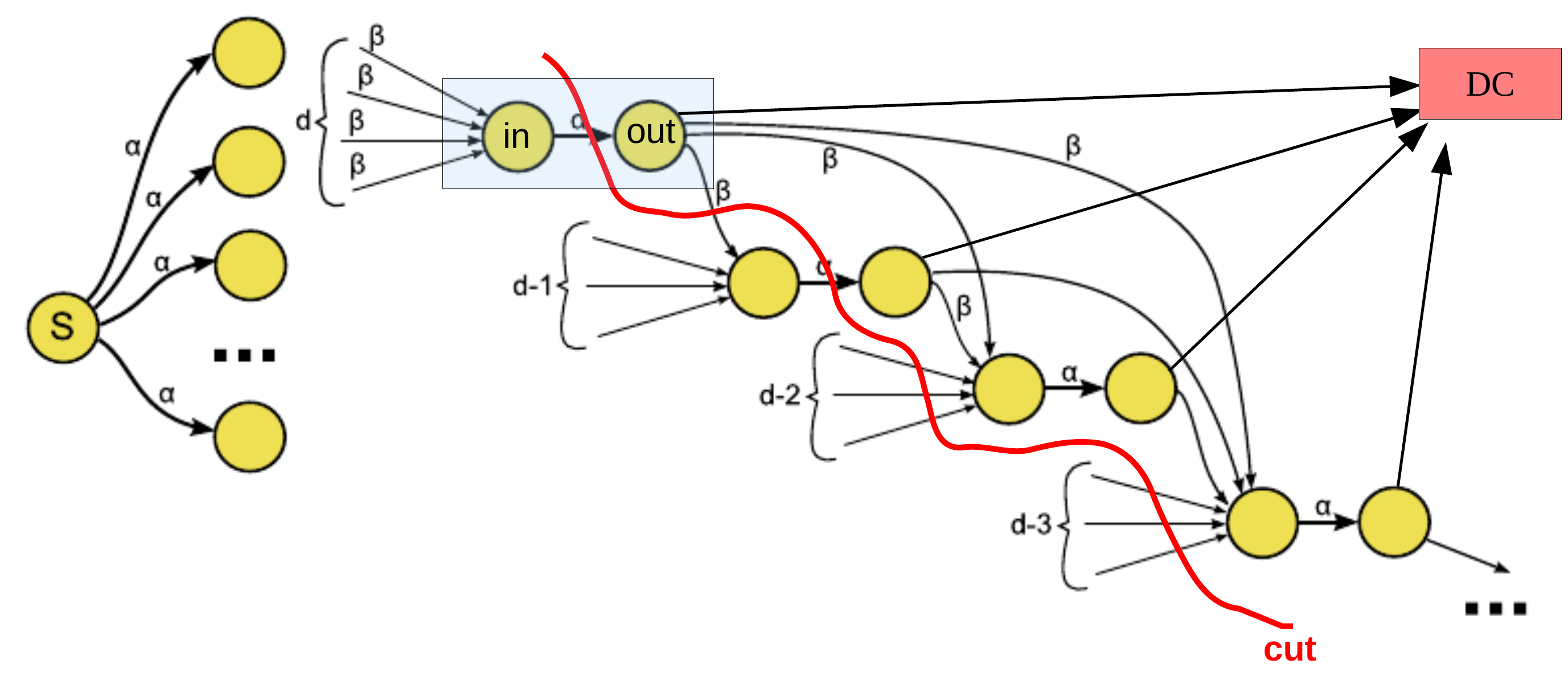}
\caption{Cut-set bound.}
\label{fig:cut_set_bound}
\end{minipage}
\end{figure}

In the regenerating-code framework, each of the $n$ nodes in the network stores $\alpha$ code symbols drawn from a finite field $\mathbb{F}_q$.  A data collector can download the data by connecting to any $k$ nodes (Fig.~\ref{fig:data_collection}) and node repair (Fig.~\ref{fig:node_repair}) is accomplished by connecting to any $d$ nodes and downloading $\beta \leq \alpha$ symbols from each node with $\alpha \leq d \beta << B$. Thus $d\beta$ is the repair bandwidth.  

Here one makes a distinction between functional and exact repair.  By functional repair, it is meant that a failed node will be replaced by a new node such that the resulting network continues to satisfy the data collection and node-repair properties defining a regenerating code.   An alternative to function repair is {\em exact repair} under which one demands that the replacement node store precisely the same content as the failed node.  From a practical perspective, exact repair is clearly preferred. 

 A cut-set bound (Fig.~\ref{fig:cut_set_bound}) based on network-coding concepts, tells us that under functional repair, given code parameters $(n,k,d, (\alpha,\beta))$, the maximum
possible size of a data file is upper bounded~\cite{DimGodWuWaiRam} by 
\bea \label{eq:cut_set_bd}
B & \leq & \sum_{i=1}^{k} \min\{\alpha,(d-i+1)\beta\} .
\eea
Furthermore, this bound has been shown to be tight as the existence of codes achieving this bound has been established using network-coding arguments related to multicasting. 

\subsection{The Storage-Repair Bandwidth Tradeoff} 

\begin{figure}[ht]
\centering
\includegraphics[height=1.5in]{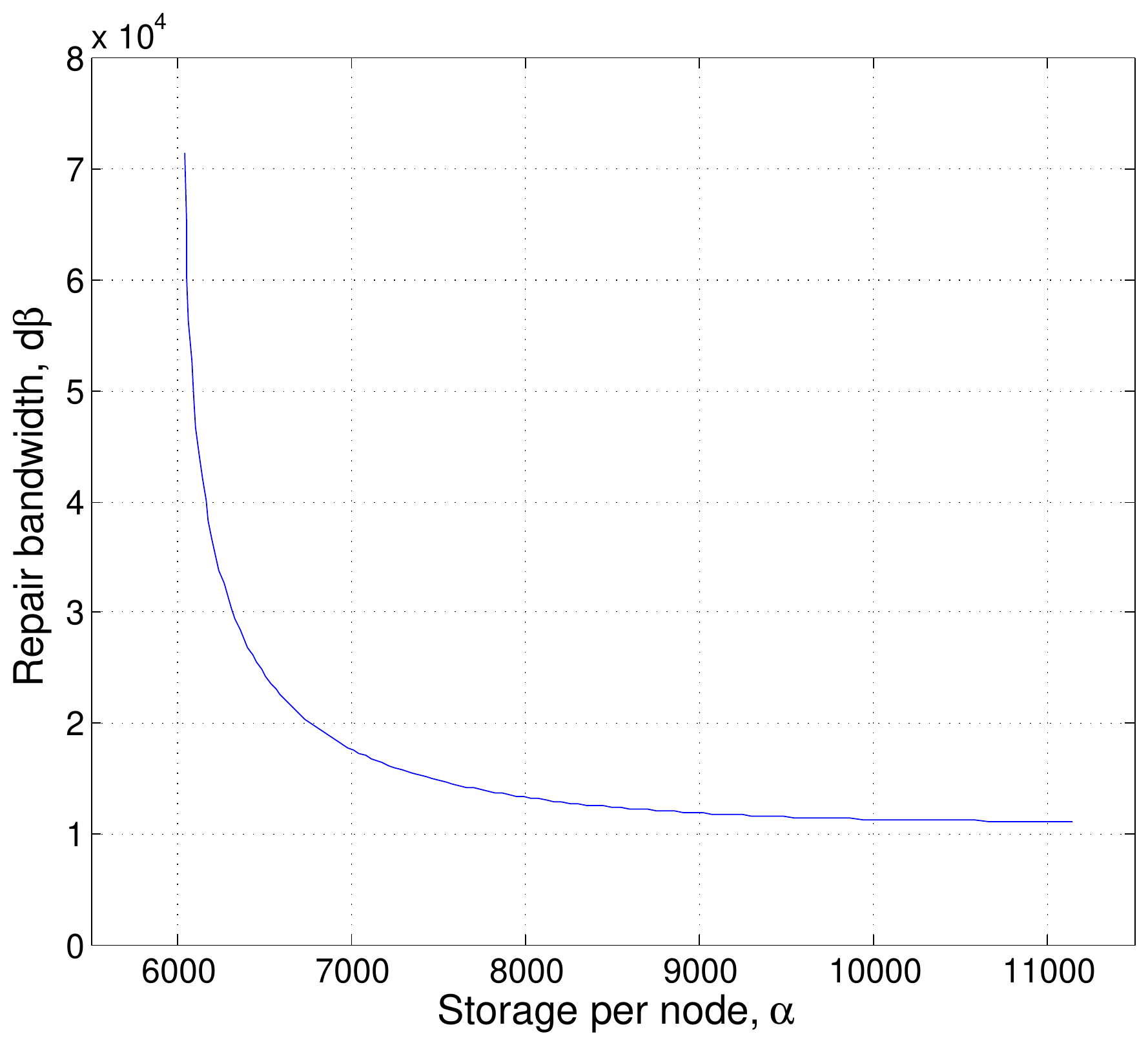}
\caption{The storage, repair-bandwidth tradeoff of regenerating codes under functional repair.}
\label{fig:tradeoff}
\end{figure}

Given $(n,k,d,B)$, there are multiple pairs $(\alpha,\beta)$ that satisfy \eqref{eq:cut_set_bd}.  This leads to the storage-repair-bandwidth (S-RB) tradeoff between $\alpha$, representing the amount of data stored and $d\beta$ representing repair bandwidth (Fig.~\ref{fig:tradeoff}).  The two extremal points in the tradeoff are respectively, the minimum-storage regenerating (MSR) and minimum bandwidth regenerating (MBR) points which correspond to the points at which the storage and repair bandwidth are respectively minimized.   The remaining points on the tradeoff curve are referred to as {\em interior points}.  


Several constructions \cite{RasShaKum_pm, CadJafMalRamSuh, PapDimCad, SuhRam, ShaRasKumRam_ia, ShaRasKumRam_rbt, TamWanBru} for regenerating codes are now available that correspond to the extreme MSR and MBR points of the tradeoff.  

\subsection{Interior Points}  

Apart from the MBR point and a small region adjacent to the MSR point, there do not exist exact-repair codes whose $(\alpha, d\beta)$ values correspond to coordinates of an interior point \cite{ShaRasKumRam_rbt} on the S-RB tradeoff under functional repair.   The possibility of approaching S-RB tradeoff asymptotically (i.e., as $B \rightarrow \infty$) using exact repair codes was recently answered in the negative in \cite{Tia}. Thus the tradeoff under exact repair has yet to be characterized. A technique known as space sharing can be used to achieve points that interpolate linearly between the MSR and MBR points. Recently a family of codes that beat the space-sharing line is proposed in the literature recently, \cite{SasKum_arxiv},\cite{SasKum_isit}, \cite{TiaAggVai_isit},\cite{TiaAggVai_arxiv}. In \cite{SasKum_arxiv}, it is also shown that these exact-repair codes achieve an interior point on the S-RB tradeoff near to the MSR point, for the case of $k=d=n-1$.

\begin{figure}[ht]
\begin{minipage}[c]{0.4\linewidth}
\centering
\includegraphics[height=1.5in]{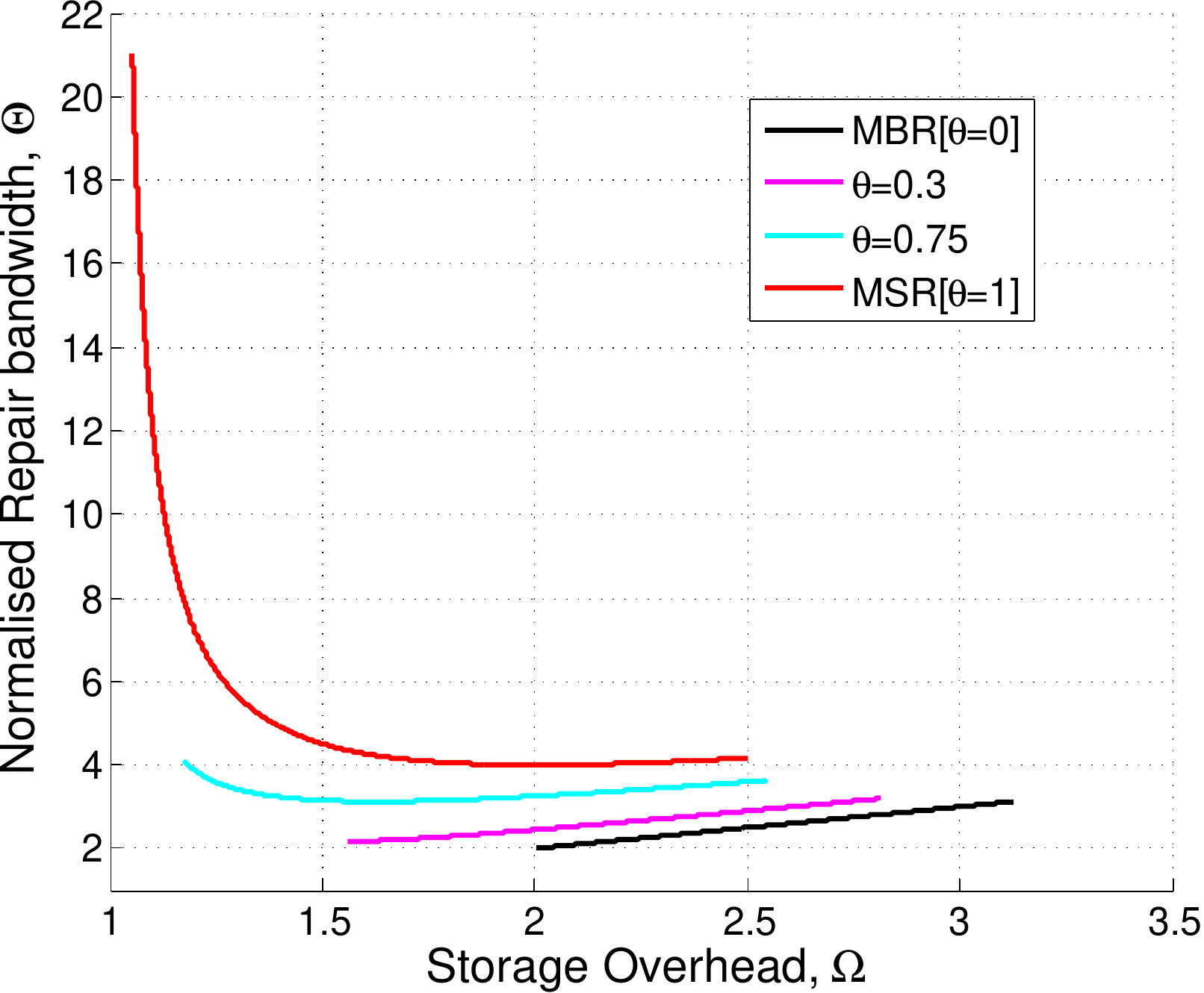}
\caption{Normalized curves.}
\label{fig:normalized_curves}
\end{minipage}
\hspace{1cm}
\begin{minipage}[c]{0.4\linewidth}
\centering
\includegraphics[height=1.5in]{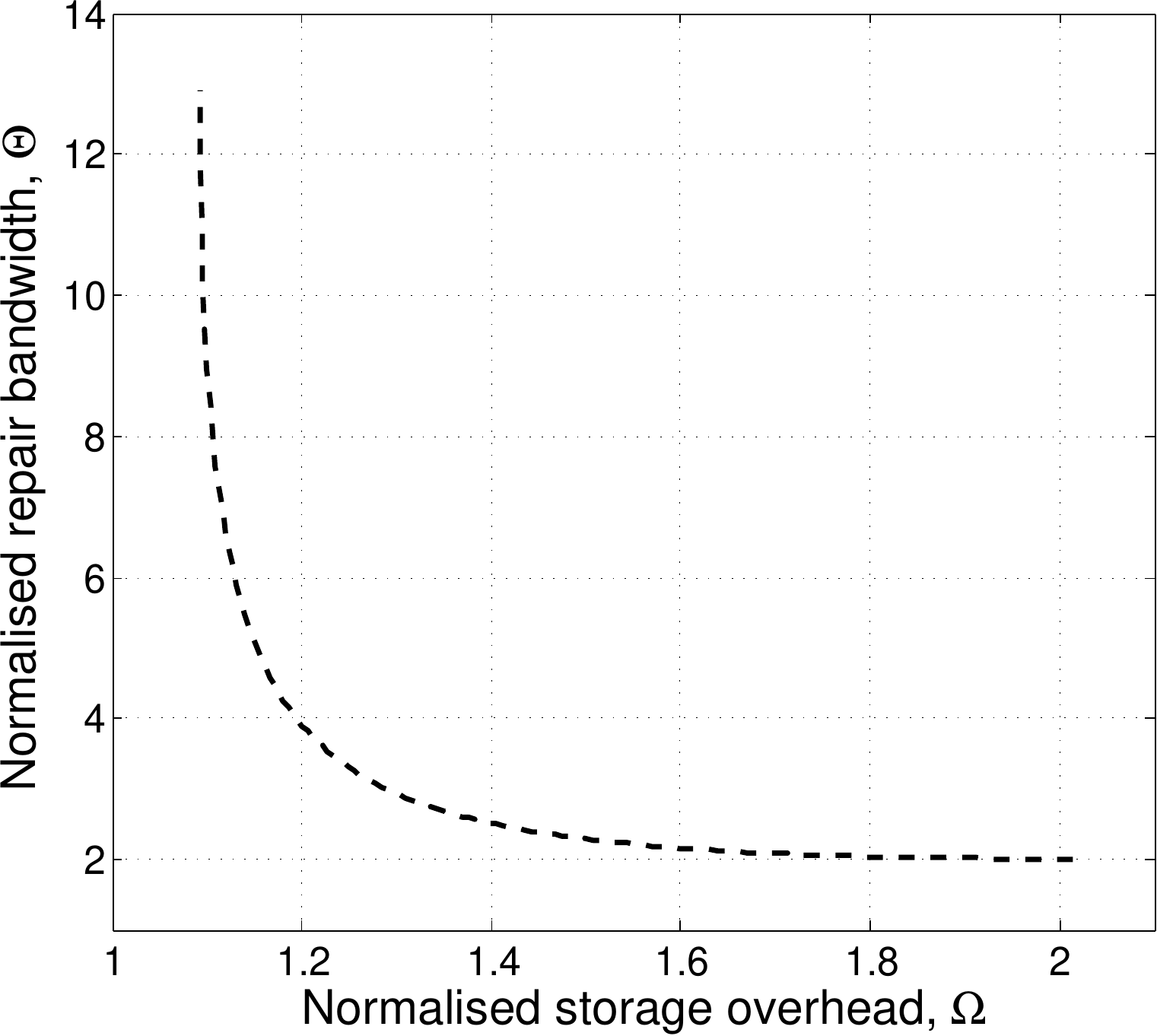}
\caption{Normalised tradeoff.}
\label{fig:normalised_tradeoff}
\end{minipage}
\end{figure}

The practical relevance of constructions for the interior points becomes clear when one considers a normalized version of the tradeoff that plots the normalized parameters $\bar{\alpha}  = \frac{n\alpha}{B}$ and $\bar{\gamma}  = \frac{nd\beta}{B}$.  The quantity $\bar{\alpha}$ represents the storage overhead while the normalization of repair bandwidth takes into account that the number of independent node failures will be proportional to $n$.   In terms of $\bar{\alpha} , \bar{\gamma}$, the cost of operating the network per unit data symbol and unit time can be expressed in the form $c_1 \bar{\alpha} + c_2 \bar{\gamma}$. The normalized tradeoff is a function of $(\frac{k}{n}, \frac{d}{n})$.  Fig.~\ref{fig:normalized_curves} plots the normalized tradeoff for the  fixed value of $\frac{d}{n}=\frac{n-1}{n}$ obtained by setting $d=(n-1)$ as $\frac{k}{n}$ is varied.  Each point on the classical tradeoff now becomes a curve as $\frac{k}{n}$ varies.  This plot shows for instance, that MBR codes can only operate in the regime where the storage overhead is $\geq 2$ and that potentially, every interior point has as important a role to play in the design of efficient regenerating codes, as do the MSR and MBR points.  Selecting the outer hull of the plots shown in Fig.~\ref{fig:normalized_curves} corresponding to a particular $\frac{k}{n}$ leads to the normalized tradeoff shown in Fig.~\ref{fig:normalised_tradeoff}.

In the present paper, we will establish a tighter tradeoff that holds in the case of regenerating codes with exact repair, for any values of $(n,k,d)$. 

\section{Bounds on the Joint Entropy of Repair Data}

It can be seen that the upper bound on file size given in \eqref{eq:cut_set_bd} is independent of the number of nodes $n$.  In this paper, we will obtain an upper bound on the size $B$ of the data file that is tighter than the bound corresponding to functional repair.  We will do this for the case $n=(d+1)$.  The resultant upper bound on file size will continue to also apply to a regenerating code having a larger value of $n$ but with the remaining parameters $\left\{(k,d), \{\alpha, \beta\}\right\}$ unchanged.  In this way, our tightening of the bound carried out for the case $n=(d+1)$ will continue to hold for the case of general $n$.  

Let $\mathcal{C}$ be a regenerating code over $\mathbb{F}_q$ having parameters $\left\{(n=(d+1),k,d), \{\alpha, \beta\}, B\right\}$. 
Let $W_i$ denote the random variable corresponding to the content of the $i$-th node, $1 \leq i \leq d+1$. Let $S_x^y$ denote the random variable corresponding to the helper data sent by helper node $x$ to the replacement node for node $y$. This is meaningful since under the assumption $n=(d+1)$, there is only one set of possible helper nodes for any failed node. From the definition of a regenerating code, it follows that 
\bean
H(W_i) & \leq & \alpha , \\
H(S_x^y) & \leq & \beta. 
\eean
Given two subset $X,Y \subseteq [n]$, we define: 
\bean
S_X^Y & =& \left\{ S_x^y \mid x \in X, y \in Y, x \neq y \right\}.
\eean
In what follows, we will use the notation $[i], 1 \leq i \leq n$ to denote $\{1, 2, \ldots, i \}$. For $1 \leq i \leq j \leq n$, $[i \ j]$ denotes the set $\{i, i+1, \ldots, j \}$. 

\begin{figure}[h!]
\begin{center}
\includegraphics[height=2.5in]{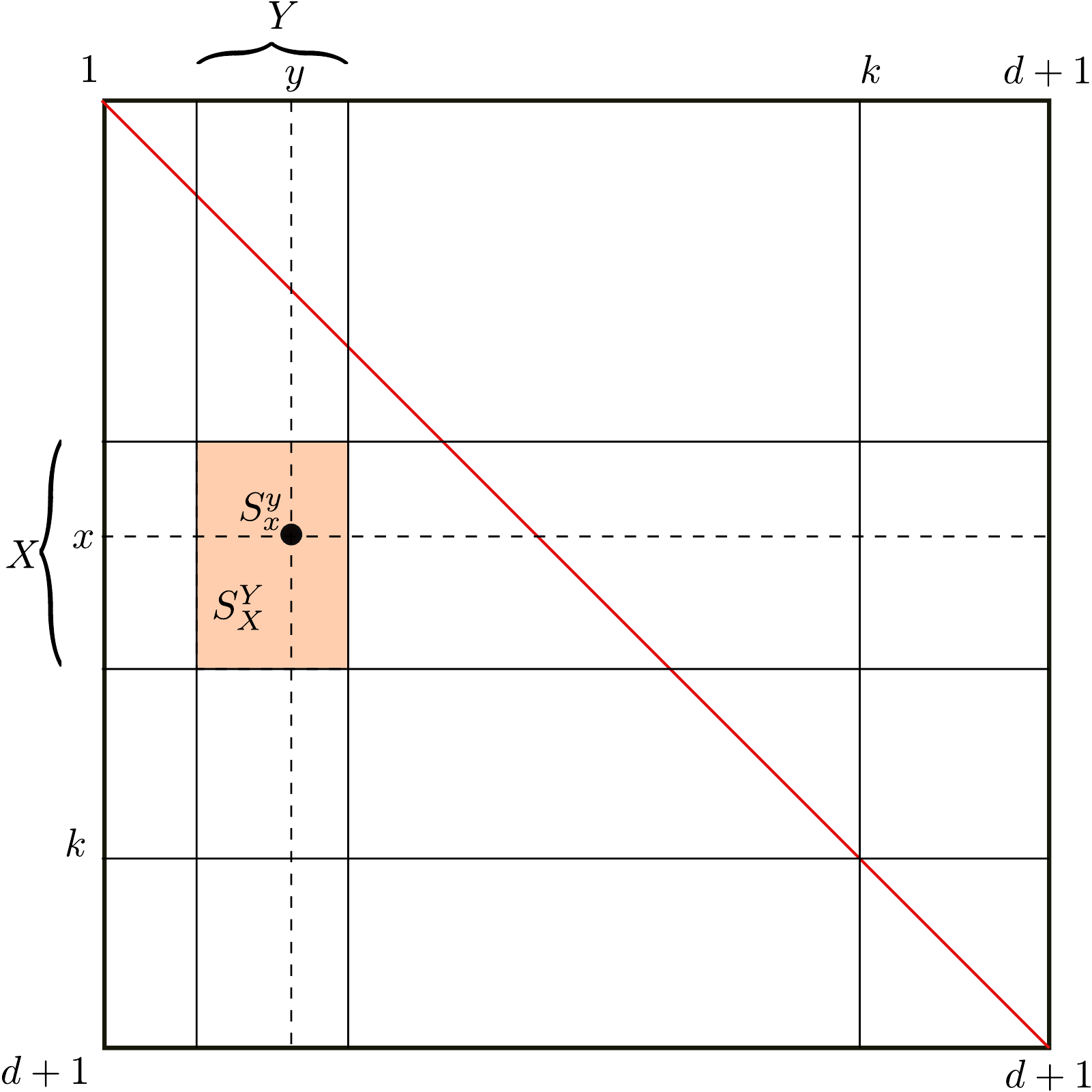}
\caption{The repair matrix.} \label{fig:repairmatrix}
\end{center} 
\end{figure}

The set of random variables $S_{[d+1]}^{[d+1]}$ can be schematically represented in a $(d+1) \times (d+1)$ array with an empty diagonal as given in Fig.~\ref{fig:repairmatrix}. In this matrix, the rows correspond to the helper nodes, and the columns correspond to the nodes getting regenerated. The point intersected by the $x$-th row and $y$-th column represents the helper data random variable $S_x^y$. This matrix will be referred to as the repair matrix. 

\paragraph{Case of an Optimal Functional Repair Code} 

We consider an optimal functional repair code ${\cal \hat{C}}$ with the same set of parameters $\left\{(n,k,d), \{\alpha, \beta\}\right\}$. Let $\hat{B}$ denote the optimal filesize for these parameters. Given any random variable or set of random variables associated with $\mathcal{C}$, we will use a hat to denote the corresponding quantity in the case of a code ${\cal \hat{C}}$ that achieves the functional repair cut-set bound with equality. 

Consider the point on the S-RB tradeoff corresponding to the assignment 
\beq
\alpha = (d-p+1) \beta - \theta ,
\eeq
with $p \in \{1,\ldots, k \}$ and $\theta \in [0,\beta)$. Here, $\theta = 0$ when $p = k$. Then the optimal functional repair file size $\hat{B}$ is known to be given by $p\alpha + \sum_{i=p+1}^{k} (d-i+1)\beta$, \cite{DimGodWuWaiRam}. Thus we have,
\bea
p\alpha + \sum_{i=p+1}^{k} (d-i+1)\beta & = & \hat{B} , \\
& = & H(\hat{W}_1, \hat{W}_2, \ldots, \hat{W}_k), \\
& = & \sum_{i=1}^{k} H(\hat{W}_i \mid \hat{W}_{[i-1]}) \label{eq:fnrepair}.
\eea

This implies that, for any set $A=[1 \ i-1]$ and $L=[i \ j], 1\leq i \leq j \leq k$,
\bea
H(\hat{W}_L \mid \hat{W}_{A}) = \sum_{\ell =i}^{j} \min \{ \alpha , (d-\ell + 1)\beta \}, \\
\text{In particular, } H(\hat{W}_i \mid \hat{W}_{[i-1]}) = \min \{ \alpha , (d-\ell + 1)\beta \} .
\eea

\paragraph{Case of an Exact-repair Code}  We consider next the exact-repair code ${\cal C}$ with file size $B$ satisfying:
\bean
B & = &  p \alpha + \sum_{i=p+1}^k (d-i+1)\beta -\epsilon.  
\eean

In the case of ${\cal C}$, clearly from the definition of regenerating codes,
\bea
H(W_i \mid W_{[i-1]}) & \leq & \min \{ \alpha , (d-i+1)\beta \} \ \ 1 \leq i \leq k \label{eq:basic1},
\eea

and for sets $A=[i-1], L = [i \ j], i \leq j \leq k$,
\bea
H(W_L \mid W_A) & \leq & \sum_{\ell =i}^{j} \min \{\alpha , (d-\ell +1) \beta \} \label{eq:basic2}.
\eea
Next, we have that: 
\bean
\sum_{i=1}^k \min\{\alpha, (d-i+1)\beta\} - \epsilon & = &  H(W_{[k]}) \\ 
 & = & H(W_A) + H(W_L \mid W_A) + H(W_{[j+1 \ k]} \mid W_A, W_L) \\
 & = & \sum_{i=1}^{k} H(W_i \mid W_{[i-1]}) \\
 & \leq & \sum_{i=1}^k \min\{\alpha, (d-i+1)\beta\} \\
 & = & \sum_{i=1}^{k} H(\hat{W}_i \mid \hat{W}_{[i-1]}) .
\eean 

Thus we obtain a lower bound on the conditional entropy of node data: 
\bea
H(W_L \mid W_{A})  & \geq & H(\hat{W}_L \mid \hat{W}_{A}) - \epsilon  \\
& = & \sum_{\ell =i}^{j} \min \{\alpha , (d-\ell +1) \beta \} - \epsilon \label{eq:basic4} \\
\text{In particular, }H(W_i \mid W_{[i-1]})  & \geq & \min \{ \alpha , (d-i+1)\beta \} - \epsilon ,\ 1 \leq i \leq k . \label{eq:basic3} 
\eea

\subsection{A Lower Bound on the Joint Entropy of Repair data} 

\begin{figure}[h!]
\begin{center}
\includegraphics[height=2.5in]{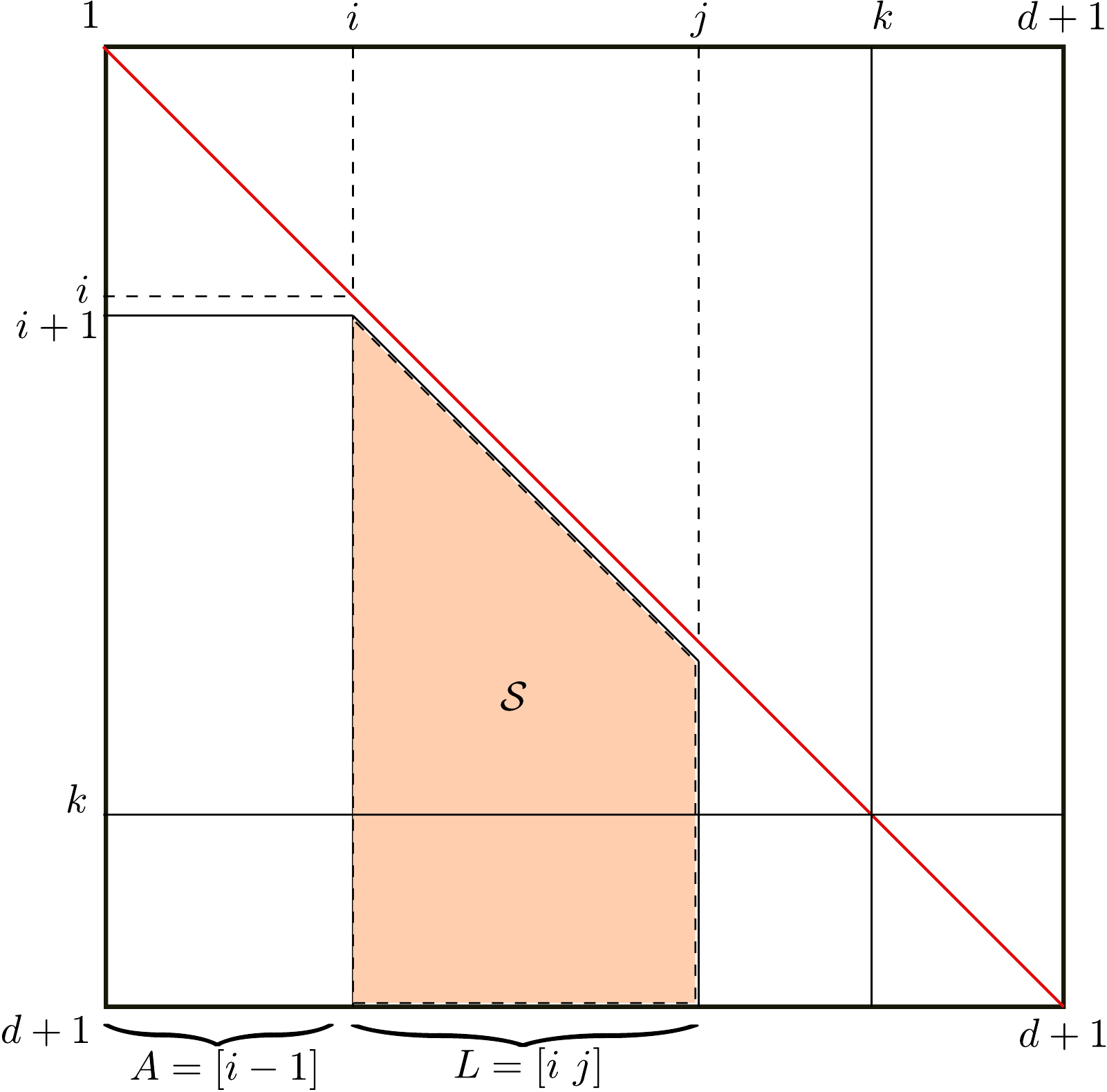}
\caption{The repair matrix.} \label{fig:repairmatrix-prop1}
\end{center} 
\end{figure}

Consider the sets $A=[i-1]$ and $L=[i \ j]$ and define
\beqn
{\cal S} = \{ S_{i+1}^i, S_{i+2}^{[i \ i+1]} , \ldots S_{j}^{[i \ j-1]} , S_{[j+1 \ d+1]}^{L}  \} .
\eeqn
The set ${\cal S}$ is pictorially represented in Fig.~\ref{fig:repairmatrix-prop1}. In the following proposition, we will get a lower bound on $H({\cal S})$. 

\bprop \label{prop:colsum} Let $L = [i \ j], i \leq j \leq k$. Then
\bea
H({\cal S} ) & \geq & \sum_{\ell = i}^{j} \min \{ \alpha, (d-\ell +1)\beta \} - \epsilon .
\eea
\eprop
\bpf Let $A=[i-1]$. Then,
\bean 
H({\cal S})  & \geq & H({\cal S} \mid W_{A}) \\
& \geq & H(W_L \mid W_{A}) \\
& \geq & H(\hat{W}_L \mid \hat{W}_{A}) - \epsilon \\
& = & \sum_{\ell = i}^{j} \min \{ \alpha, (d-\ell +1)\beta \} - \epsilon .
\eean
\epf

\subsection{An Upper Bound on the Joint Entropy of Elements within a Row} 

In this subsection, we obtain an upperbound on the joint entropy of repair data on a single row of the repair matrix. 

\bprop \label{prop:rowsum} Let $L \subset [k], |L| = \ell, m \notin L$.  Then the following inequalities hold.
\bea
H(S_m^L) & \leq & \beta + (\ell -1)\theta + \ell \epsilon \ \ , \ \ 2 \leq \ell \leq p < k \label{eq:rowsum1} \\
H(S_m^L) & \leq & 2\beta - \theta + \ell \epsilon \ \ , \ \  2 \leq \ell \leq p+1 < k \label{eq:rowsum2} 
\eea
\eprop

\bpf The proof builds on the techniques introduced in \cite{ShaRasKumRam_rbt}. Consider a set $R$ such that $ L \subseteq R \subset [k]$, $m \notin R$ and $|R| = r < k$. Then,
\bea
H(S_m^L) & = & H(S_m^L \mid W_R) + I(S_m^L : W_R) , \\
& \leq & \ell H(S_m^{j_0} \mid W_R) + I(S_m^L : W_R) ,  
\eea
where $j_0$ is such that $H(S_m^{j} \mid W_R), j \in L$ is maximum when $j = j_0$. Define $R_* = R \setminus \{j_0\}$.  We then have the series of inequalities: 
\bea
H(S_m^L) & \leq & \ell H(S_m^{j_0} \mid W_R) + I(S_m^L : W_R) \nonumber \\
& \leq & \ell \left\{ H(S_m^{j_0}) - I(S_m^{j_0} : W_R) \right\} + I(W_m : W_R) \nonumber \\
& \leq & \ell \left\{ \beta - [ H(W_R) - H(W_R \mid S_m^{j_0}) ] \right\} + I(W_m : W_R) \nonumber \\
& = & \ell \left\{ \beta - [ H(W_{R_*} ) + H(W_{j_0} \mid W_{R_*} ) - H(W_{R_*} \mid S_m^{j_0}) - H(W_{j_0} \mid S_m^{j_0}, W_{R_*} ) ] \right\} + I(W_m : W_R) \nonumber \\
& \leq & \ell \left\{ \beta - [ H(W_{R_*} ) + H(W_{j_0} \mid W_{R_*} ) - H(W_{R_*}) - H(W_{j_0} \mid S_m^{j_0}, W_{R_*} ) ] \right\} + I(W_m : W_R) \nonumber \\
& = & \ell \left\{ \beta - [ H(W_{j_0} \mid W_{R_*} ) - H(W_{j_0} \mid S_m^{j_0}, W_{R_*} ) ] \right\} + I(W_m : W_R) \nonumber \\
& = & \ell \left\{ \beta + H(W_{j_0} \mid S_m^{j_0}, W_{R_*} ) \right\} - \ell H(W_{j_0} \mid W_{R_*} ) + H(W_m) + H(W_R) - H(W_m,W_R) \nonumber \\
& \leq & \ell \left\{ \beta + H(W_{j_0} \mid S_m^{j_0}, S_{R_*}^{j_0} ) \right\} + H(W_m) + H(W_{R_*}) - (\ell - 1) H(W_{j_0} \mid W_{R_*} )  - H(W_m,W_R) \nonumber \\
\label{eq:last} & \leq & \ell (d-r+1)\beta + r \alpha - (\ell -1)H(W_{j_0} \mid W_{R_*} ) - H(W_m,W_R) 
\eea

In \eqref{eq:last}, by choosing $r=p$ and applying \eqref{eq:basic4}, we prove the first inequality in \eqref{eq:rowsum1} for $2 \leq \ell \leq r=p$. Similarly, by choosing $r=p+1$ and applying \eqref{eq:basic4}, we prove the second inequality in \eqref{eq:rowsum2} for $2 \leq \ell \leq r=p+1$.
 
\epf

\section{The Improved Tradeoff} 

In this section, we make use of Prop.~\ref{prop:colsum}, and Prop.~\ref{prop:rowsum} to derive an improved bound on the S-RB tradeoff for exact-repair regenerating codes.

\bthm \label{thm:main} In the case of any exact-repair regenerating code ${\cal C}$ with $k \geq 3$, the following tighter (in comparison with the file size under functional repair) upper bound on the tradeoff between $\alpha$ and $d\beta$ is characterized by:
\ben
\item For $p=1, \ \theta \neq 0$,
\bean
B & \leq & \hat{B} - \epsilon_1
\eean
\item For $p \in \{ 2, 3, \ldots, k-2 \}$,
\bean
B & \leq & \hat{B} - \max \{ \epsilon_0, \epsilon_1 \}
\eean
\item For $p=k-1, \ \theta < \left(\frac{d-k+1}{d-k+2}\right)\beta$,
\bean
B & \leq & \hat{B} - \epsilon_0 
\eean
\een

where $\epsilon_0$ and $\epsilon_1$ are as given in Tab.~\ref{tab:eps_table}.

\ethm

\begin{table}[h]
\centering

\begin{tabular}{||c|c||}  \hline
\hline
 &   \\
Regime of $(p,\theta)$ & Lower bounds $\epsilon_0$ , $\epsilon_1$  on $\epsilon = \hat{B}-B$ \\
 &   \\
\hline
\hline
 &   \\
$ \begin{array}{c} p \in \{ 2, 3, \ldots, k-1 \} \text{ for all } \theta \\
\text{For } p=k-1, \ \theta < \frac{d-k+1}{d-k+2}\beta
\end{array}$ & 
\large
$\begin{array}{lcl}
& & \text{Let } q_0 = \left\lfloor \frac{k-p+1}{p} \right\rfloor  \\
&& \\
\epsilon_0 & = & \left\{ \begin{array}{lc} \frac{(d-k+1)(k-p)(\beta - \theta) \ - \ \theta}{(d-k+1)(k-p+1) \ + \ 1}, & k-p+1 < p. \\
& \\
\frac{\left(d-\frac{p(q_0+3)}{2}+2 \right)q_0(p-1)(\beta - \theta) \ - \ \theta}{\left(d-\frac{p(q_0+3)}{2}+2\right)q_0p \ + \ 1}, & k-p+1 \geq p . \end{array} \right. \end{array}$ \\
 &   \\
\hline
 &   \\
$ \begin{array}{c} p \in \{ 1, 2, \ldots, k-2 \} \text{ for all } \theta \\
\text{For } p=1, \ \theta \neq 0
\end{array}$ & 
\large
$\begin{array}{lcl}
& & \text{Let } q_1 = \left\lfloor \frac{k-p}{p+1} \right\rfloor  \\
&& \\
\epsilon_1 & = & \left\{ \begin{array}{lc} \frac{(d-k+1)\left[(k-p-2)\beta \ + \ \theta\right]}{(d-k+1)(k-p) \ + \ 1}, & k-p < p+1. \\
& \\
\frac{\left(d-\frac{(p+1)(q_1+3)}{2}+2 \right)q_1 \left[(p-1)\beta \ + \ \theta\right] }{\left(d-\frac{(p+1)(q_1+3)}{2}+2\right)q_1(p+1) \ + \ 1}, & k-p \geq p+1. \end{array} \right. \end{array}$ \\
 &   \\
\hline \hline
\end{tabular}
\caption{Lower Bounds on the quantity $\hat{B}-B$}
\label{tab:eps_table}
\end{table}

\bcor The optimal tradeoff between $\alpha$ and $d\beta$ for any exact-repair regenerating code, with $k \geq 3$, for a fixed filesize $B$ is strictly away from that of functional-repair regenerating codes whenever $(p = 1, \theta \neq 0)$, $p \in \{2,3,\ldots , k-2\}$ or $(p=k-1, \ \theta < \frac{d-k+1}{d-k+2}\beta)$.
\ecor
\bpf Let
\bean
\delta = \left\{ \begin{array}{lc} \epsilon_1 & p = 1, \theta \neq 0 \\
							\max \{\epsilon_0, \epsilon_1\} & p \in \{2,3,\ldots , k-2 \} \\
							\epsilon_0 & p = k-1, \theta < \frac{d-k+1}{d-k+2}\beta \end{array} \right. 
\eean

Let $\alpha$ be related to $\beta$ as $\alpha = (d-p+1)\beta - \theta=(d-p+1)\beta - t\cdot\beta, \ \ t \in [0,1)$ by a fixed pair $(p,t)$ that falls in the range given. Then for a code with the file size $B$,
\bean
\frac{\beta}{B} & \geq & \frac{\beta}{\hat{B} - \delta }, \ \ \ \text{(using Thm.~\ref{thm:main})}\\
& = & \frac{\beta}{\hat{B}} \cdot \frac{1}{1-\left(\frac{\delta}{\hat{B}} \right)} \\
& = & \frac{\beta}{\hat{B}} \cdot \frac{1}{ 1 - \left(\frac{\delta}{ \beta \sum_{i=1}^{k} \min \{ (d-p+1)-t , (d-i+1) \} } \right)}  \\
& \geq & \frac{\beta}{\hat{B}} + \delta_0, 
\eean
for some $\delta_0 > 0$, determined by the constants $\frac{\epsilon_0}{\beta}$ and $\frac{\epsilon_1}{\beta}$. Futher, it can be seen that $\frac{\epsilon_0}{\beta}$ and $\frac{\epsilon_1}{\beta}$ are independent of $\beta, B$ and dependent only on the fixed values of $p,t,k$ and $d$. Hence the proof.
\epf

\begin{note} When $p=k-1$ (i.e., the region close to the MSR point), the new outerbound is strictly away from the fuctional repair tradeoff when $0 \leq \theta < \left(\frac{d-k+1}{d-k+2}\right)\beta$ . This range of $\theta$ coincides with the range for $\theta$ for which authors of \cite{ShaRasKumRam_rbt} proved non-existence of exact-repair codes operating at functional repair tradeoff.
\end{note}

\begin{example} $(n,k,d)=(4,3,3)$

In this case, we need to consider the cases of $p = 1, \ 0 < \theta < \beta$ and $p=2, \ 0 \leq \theta < \frac{\beta}{2}$. For every $(p,\theta)$, we have $\alpha = (4-p)\beta - \theta$ .

When $p = 1, \ 0 < \theta < \beta$,
\bea
\epsilon_1 & = & \frac{\theta}{3} \ = \ \frac{3\beta - \alpha}{3}. \\
\text{Then we have, } B & \leq & \hat{B} -  \epsilon_1 \\
\text{leading to, } 3B & \leq & 4\alpha + 6\beta , \ \ p = 1, \ 0 < \theta < \beta . \label{eq:433_1}
\eea
When $p=k-1=2, \ 0 \leq \theta < \frac{\beta}{2}$,
\bea
\epsilon_0 & = & \frac{\beta - 2\theta}{3} \ = \ \frac{\beta - 2(2\beta - \alpha)}{3}. \\
\text{Then we have, } B & \leq & \hat{B} -  \epsilon_0 \\ 
\text{leading to, } 3B & \leq & 4\alpha + 6\beta, \ \ p=2, \ 0 \leq \theta < \frac{\beta}{2}. \label{eq:433_2} 
\eea
Equations \eqref{eq:433_1} and \eqref{eq:433_2} characterize the new outerbound. Remarkably, the bound coincides with the optimal tradeoff, proved in \cite{Tia}. See Fig.~\ref{fig:433}.

\begin{figure}[h!]
\centering
\includegraphics[height=3in]{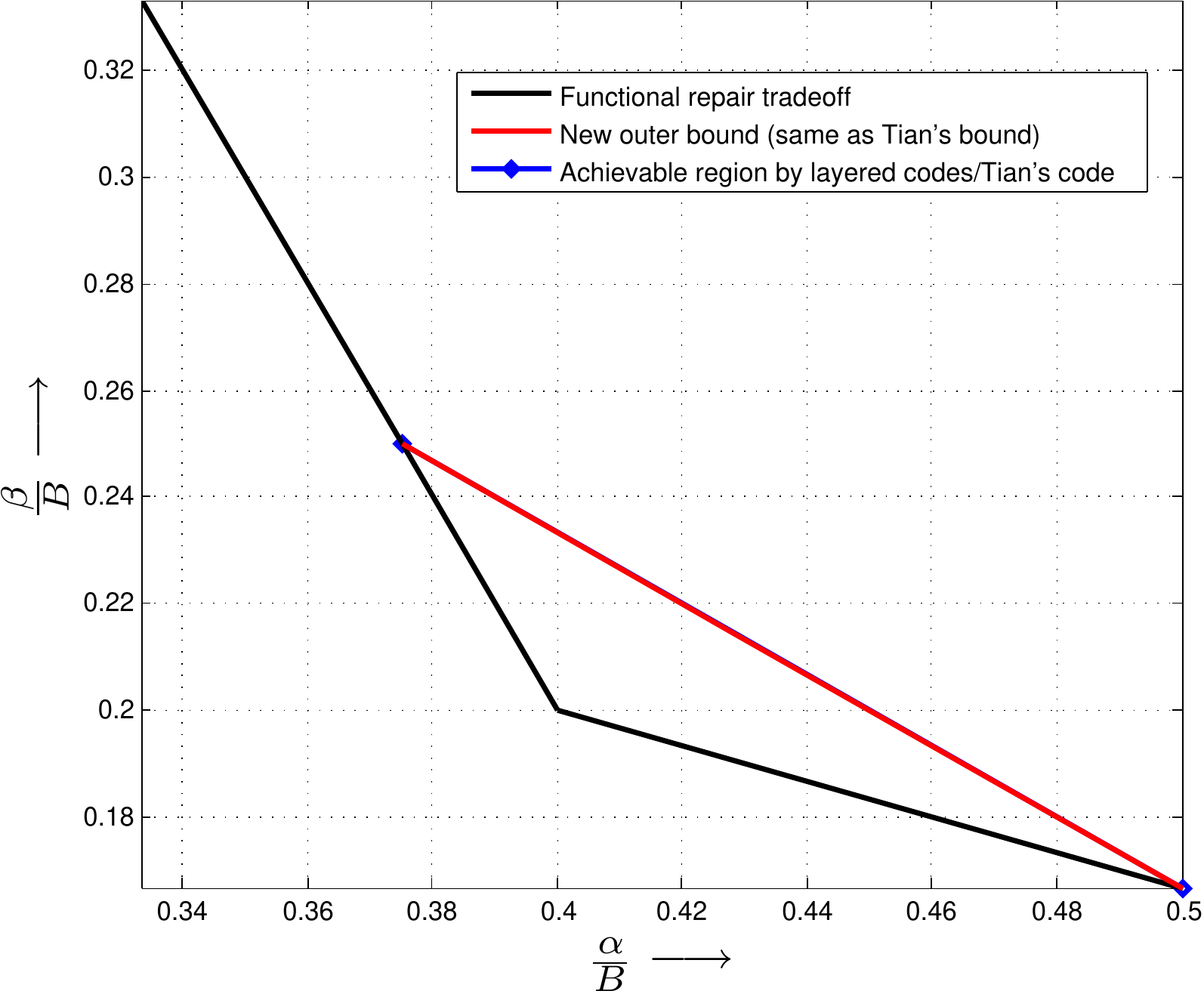}
\caption{The new outer bound coincides with the optimal tradeoff for $(4,3,3)$, shown in \cite{Tia}.} \label{fig:433}
\end{figure}

\end{example}

\begin{example} $(n,k,d)=(5,4,4)$

In this case, we need to consider the cases of $p = 1$ with $\ 0 < \theta < \beta$, $p=2$ with $\ 0 \leq \theta < \beta$, and $p=3$ with $\ 0 < \theta < \frac{\beta}{2}$. For every $(p,\theta)$, we have $\alpha = (5-p)\beta - \theta$.

When $p = 1, \ 0 < \theta < \beta$,
\bea
\epsilon_1 & = & \frac{2\theta}{5} \ =  \ \frac{2(4\beta - \alpha)}{5} \\
\text{Then we have, } \ \ B & \leq & \hat{B} -  \epsilon_1  \\
\text{leading to, } \ \ 5B & \leq & 7\alpha + 22\beta, \ \ p = 1, \ 0 < \theta < \beta . \label{eq:544_1}
\eea
When $p=2, \ 0 \leq \theta < \beta$,
\bea
\epsilon_1 & = & \frac{\theta}{3} \ = \ \frac{3\beta - \alpha}{3} \\
\epsilon_0 & = & \frac{2\beta - 3\theta}{5} \ = \ \frac{2\beta - 3(3\beta - \alpha)}{5} \\
\text{Then we have, } \ \ B & \leq & \hat{B} - \max \{\epsilon_0 ,\epsilon_1\} \\
\text{leading to, } \ \ 5B & \leq & 7\alpha + 22\beta , \ \ \ \frac{1}{3} \geq \frac{\alpha}{\beta} > \frac{18}{7} \label{eq:544_2} \\
3B & \leq & 7\alpha + 6\beta , \ \ \ \frac{18}{7} \geq \frac{\alpha}{\beta} > \frac{1}{2} \label{eq:544_3}
\eea
When $p=k-1=3, \ 0 \leq \theta < \frac{\beta}{2}$,
\bea
\epsilon_0 & = & \frac{\beta - 2\theta}{3} \ = \ \frac{\beta - 2(2\beta - \alpha)}{3} \\
\text{Then we have, } \ \ B & \leq & \hat{B} -  \epsilon_0 \\ 
\text{leading to, }  \ \ 3B & \leq & 7\alpha + 6\beta , \ \ p=3, \ 0 \leq \theta < \frac{\beta}{2}\label{eq:544_4}
\eea
Equations \eqref{eq:544_1}, \eqref{eq:544_2}, \eqref{eq:544_3} and \eqref{eq:544_4} characterize the new outerbound. In Fig.~\ref{fig:544}, the bound is plotted against the region achievable by the layered codes, \cite{SasKum_isit}, \cite{TiaAggVai_isit}. When $\alpha = 2\beta$, the bound is achieved by the layered code. Thus for the case of $(5,4,4)$ system, the optimal tradeoff point is characterized when $\alpha= 2\beta$.

\begin{figure}[h!]
\centering
\includegraphics[height=3in]{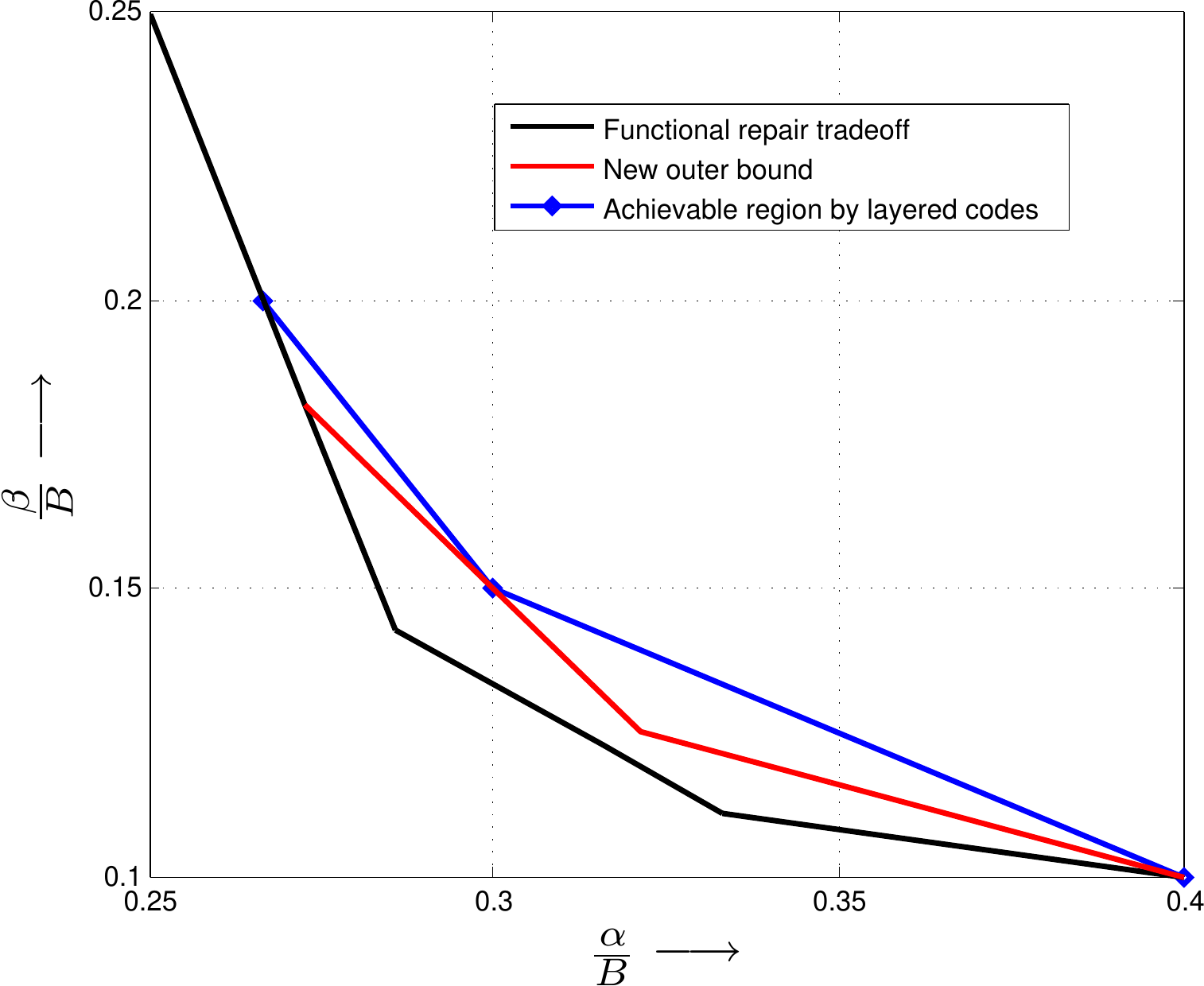}
\caption{The new outer bound is compared against the achievable region of layered codes for $(5,4,4)$.} \label{fig:544} 
\end{figure}

\vspace{0.1in}

\bpf[Proof of Theorem \ref{thm:main}] The method of the proof is to derive lower bounds on $\epsilon = \hat{B} - B$ in various cases. Towards this, we will first consider an appropriately chosen collection of repair data random variables. The joint entropy of this collection of random variables is bounded below and above respectively invoking Prop.~\ref{prop:colsum} and Prop.~\ref{prop:rowsum}. The resulting inequality leads to a lower bound for $\epsilon$.

\vspace{0.1in}

{\em Case 1:} $p \in \{2,3,\ldots, k-1\}$

We set $q_0=\left\lfloor \frac{k-p+1}{p} \right\rfloor$. We will have two subcases for $q_0 \geq 1$ and $q_0=0$. 

\vspace{0.1in}

{\em Case 1(a):} $q_0 \geq 1$

\begin{figure}[h!]
\begin{center}
\includegraphics[width=2.9in]{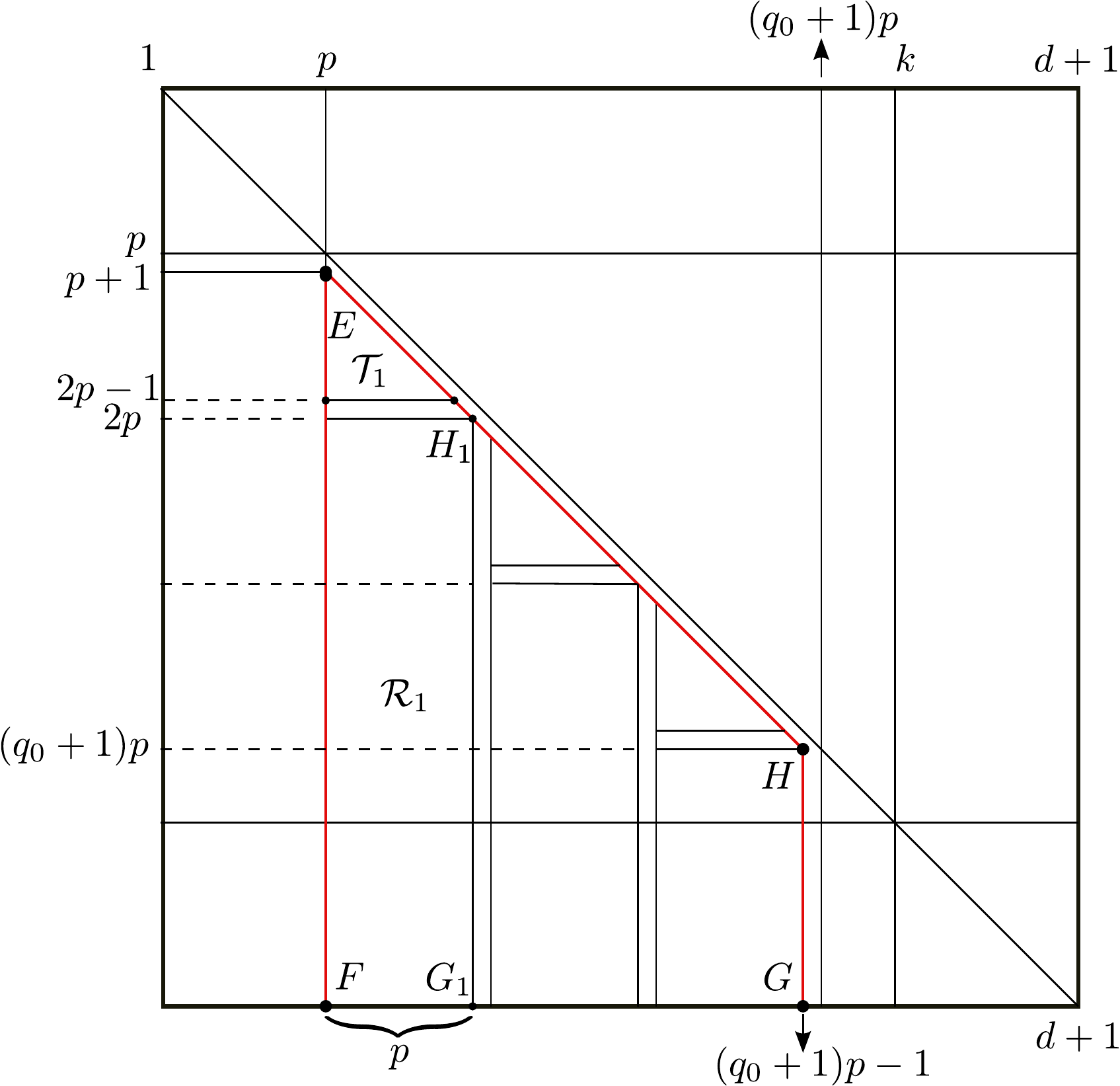}
\caption{The trapezium $EFGH$ marked in red in the repair matrix indicates the collection of random variables, ${\cal S}$, considered in {\em Case 1(a)}. Here $q_0=\left\lfloor \frac{k-p+1}{p} \right\rfloor$. } \label{fig:case1a}
\end{center}
\end{figure}


We define a set of random variables ${\cal S} = \{ S_x^y \mid p+1 \leq x \leq d+1, p \leq y \leq (q_0+1)p-1 , x > y \}$ and it corresponds to the trapezium $EFGH$ in the repair matrix, as shown in Fig.~\ref{fig:case1a}. These random variables are associated with the helper data for the set of nodes $Y = \{p, p+1, \ldots , (q_0+1)p-1\}$. We split $Y$ into $q_0$ groups of $p$ nodes each in order, and the correponding subsets of ${\cal S}$ are denoted by ${\cal S}_i, i = 1, 2, \ldots, q_0$. Pictorially, ${\cal S}_1$ is associated with the trapezium $EFG_1H_1$ in Fig.~\ref{fig:case1a}. Similarly every ${\cal S}_i$ is associated with a smaller trapezium contained within $EFGH$. The set ${\cal S}_i$ can be again viewed as the union of two subsets ${\cal R}_i$ and ${\cal T}_i$, respectively associated with the largest rectangle within the trapezium, and the remaining triangular region. These sets are formally defined as,
\bean
{\cal S}_i & = & \{ S_x^y \mid S_x^y \in {\cal S}, pi \leq y \leq p(i+1)-1 \}, \ \ i = 1,2,\ldots , q_0 \\
{\cal R}_i & = & \{ S_x^y \mid S_x^y \in {\cal S}_i, p(i+1) \leq x \leq d+1 \}, \ \ i = 1,2,\ldots , q_0 \\
{\cal T}_i & = & \{ S_x^y \mid S_x^y \in {\cal S}_i, pi+1 \leq x \leq p(i+1)-1 \}, \ \ i = 1,2,\ldots , q_0 
\eean
Note that ${\cal S}_i = {\cal R}_i \cup {\cal T}_i$. Now we proceed towards bounding the joint entropy $H({\cal S})$. We have,
\bea
H({\cal S}) &  \leq & \sum_{i=1}^{q_0} H({\cal R}_i) + \sum_{i=1}^{q_0} H({\cal T}_i) \\
& \leq & \sum_{i=1}^{q_0} \left(d-(i+1)p+2\right) \cdot  [\beta + (p-1)\theta + p\epsilon] + \sum_{i=1}^{q_0} \frac{p(p-1)\beta}{2} \label{eq:upper1a}
\eea
In the second inequality, we use \eqref{eq:rowsum1} of Prop.~\ref{prop:rowsum} to obtain the upper bound on $H({\cal R}_i)$. On the other hand, using Prop.~\ref{prop:colsum}, we also have,
\bea
H({\cal S}) &  \geq &  \sum_{i=p}^{(q_0+1)p-1} \min \{\alpha , (d-i+1)\beta \} - \epsilon \\
& = & \left[ \sum_{i=p}^{(q_0+1)p-1} (d-i+1)\beta \right] - \theta - \epsilon \label{eq:lower1a}
\eea
Matching the bounds in \eqref{eq:upper1a} and \eqref{eq:lower1a}, we must have 
\bea
\epsilon & \geq & \frac{\left(d-\frac{p(q_0+3)}{2}+2 \right)q_0(p-1)(\beta - \theta) \ - \ \theta}{\left(d-\frac{p(q_0+3)}{2}+2\right)q_0p \ + \ 1} \label{eq:eps1a}
\eea

\vspace{0.1in}

{\em Case 1(b):} $q_0=0$

\begin{figure}[h!]
\begin{center}
 \includegraphics[width=2.8in]{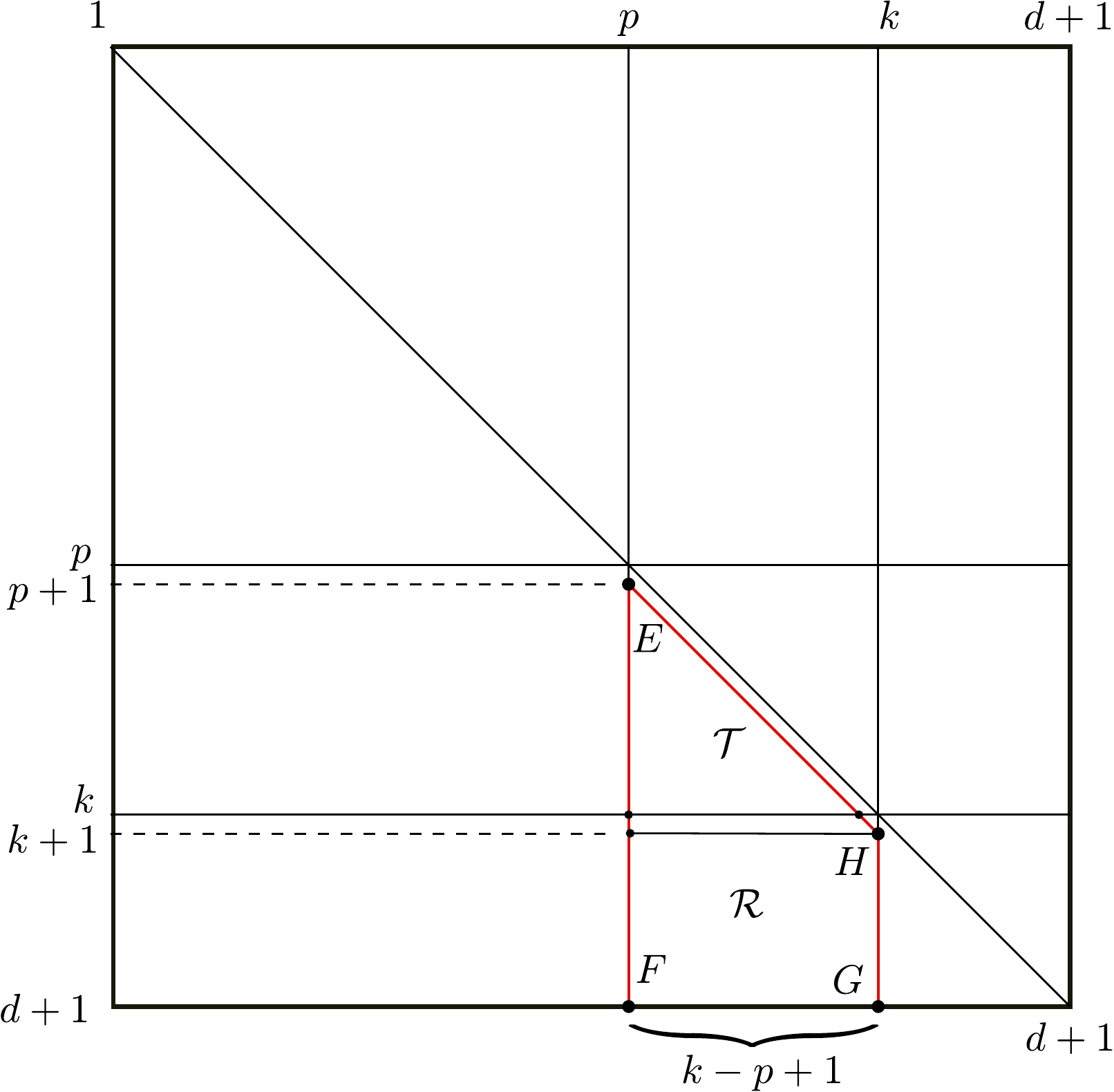}
\caption{The trapezium $EFGH$ marked in red in the repair matrix indicates the collection of random variables, ${\cal S}$, considered in {\em Case 1(b)}. } \label{fig:case1b}
\end{center}
\end{figure}

In this case, we define 
\bean
{\cal S} & = & \{ S_x^y \mid p+1 \leq x \leq d+1, p \leq y \leq k , x > y \} \\
{\cal R} & = & \{ S_x^y \mid S_x^y \in {\cal S}, k+1 \leq x \leq d+1 \}, \\
{\cal T} & = & \{ S_x^y \mid S_x^y \in {\cal S}, p+1 \leq x \leq k \} 
\eean
Note that ${\cal S} = {\cal R} \cup {\cal T}$ . In Fig.~\ref{fig:case1b}, ${\cal S}$ is represented by the trapezium $EFGH$. The largest rectangle within $EFGH$ is associated with ${\cal R}$ and the remaining triangular portion with ${\cal T}$. In a similar fashion as in {\em Case 1(a)}, we will bound the joint entropy $H({\cal S})$. Using Prop.~\ref{prop:rowsum}, we have
\bea
H({\cal S}) &  \leq & H({\cal R}) + H({\cal T}) \\
& \leq & (d-k+1) \cdot [\beta + (k-p)\theta + (k-p+1)\epsilon ] + \frac{(k-p)(k-p+1)\beta}{2} . \label{eq:upper1b}
\eea
On the other hand, using Prop.~\ref{prop:colsum},
\bea
H({\cal S}) &  \geq &  \sum_{i=p}^{k} \min \{\alpha , (d-i+1)\beta \} - \epsilon \\
& = & \left[ \sum_{i=p}^{k} (d-i+1)\beta \right] - \theta - \epsilon \label{eq:lower1b}
\eea
Matching the bounds in \eqref{eq:upper1b} and \eqref{eq:lower1b}, we must have
\bea
\epsilon & \geq & \frac{(d-k+1)(k-p)(\beta - \theta) \ - \ \theta}{(d-k+1)(k-p+1) \ + \ 1} \label{eq:eps1b}
\eea

\vspace{0.1in}

{\em Case 2:} $p \in \{1,2,\ldots, k-2\}$

We set $q_1=\left\lfloor \frac{k-p}{p+1} \right\rfloor$. We will have two subcases for $q_1 \geq 1$ and $q_1=0$. The difference in {\em Case 2} compared against {\em Case 1} lies in the choice of set of random variables ${\cal S}$ and the way we split it into subsets.

\vspace{0.1in}

{\em Case 2(a):} $q_1 \geq 1$

\begin{figure}[h!]
\begin{center}
\includegraphics[width=3.1in]{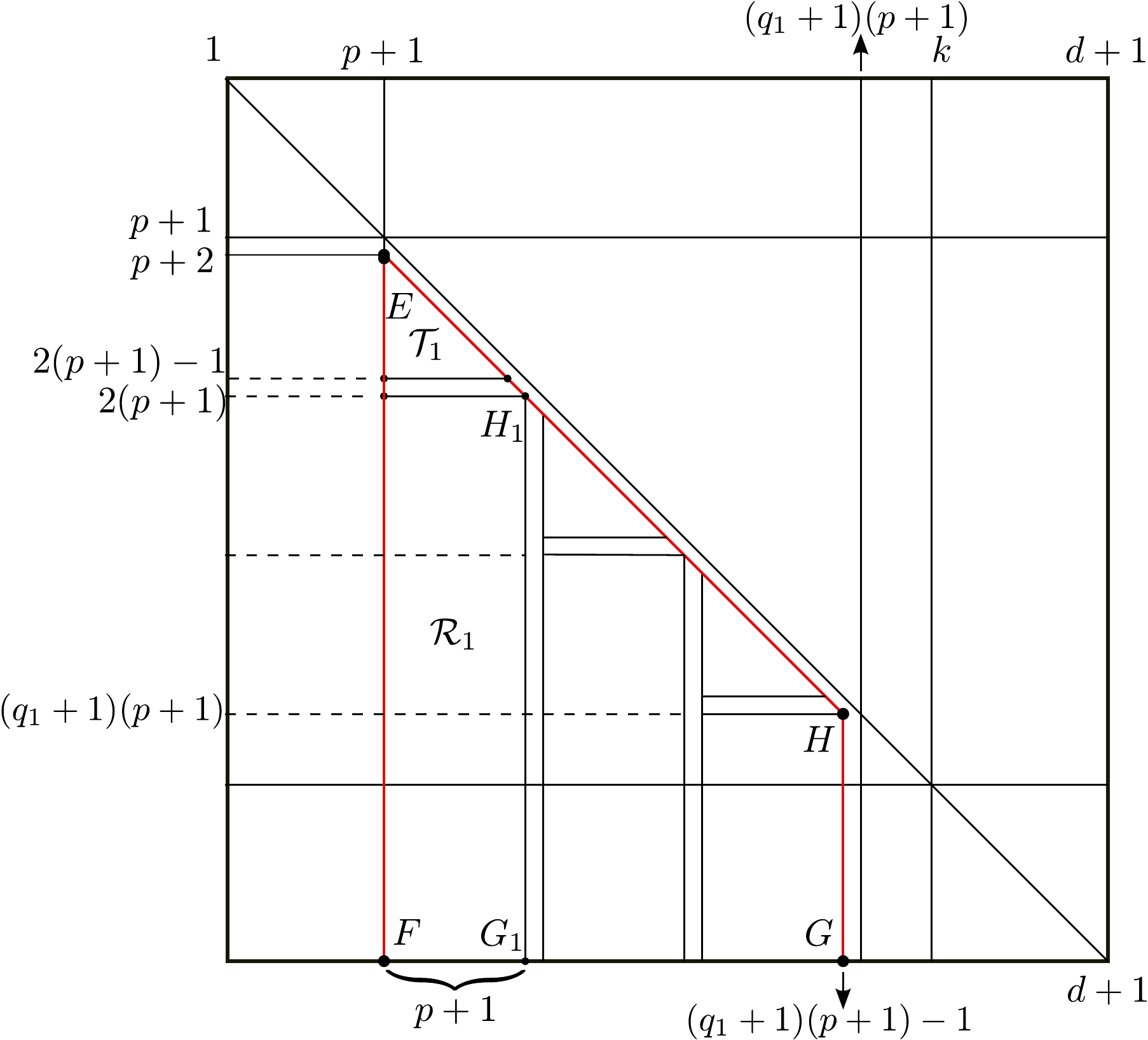}
\caption{The trapezium $EFGH$ marked in red in the repair matrix indicates the collection of random variables, ${\cal S}$, considered in {\em Case 2(a)}. Here $q_1=\left\lfloor \frac{k-p}{p+1} \right\rfloor$. } \label{fig:case2a}
\end{center}
\end{figure}


We define ${\cal S} = \{ S_x^y \mid p+2 \leq x \leq d+1, p+1 \leq y \leq (q_1+1)(p+1)-1 , x > y \}$ and it corresponds to the trapezium $EFGH$ in the repair matrix, as shown in Fig.~\ref{fig:case2a}. These random variables are associated with the helper data for the set of nodes $Y = \{p+1, p+2, \ldots , (q_1+1)(p+1)-1\}$. We split $Y$ into $q$ groups of $(p+1)$ nodes each in order, and the correponding subsets of ${\cal S}$ are denoted by ${\cal S}_i, i = 1, 2, \ldots, q_1$. A pictorial representation of how the set ${\cal S}$ is split into subsets in given in Fig.~\ref{fig:case2a}, and it is quite similar to what we had in the previous case. Similar to the previous case, we have subsets ${\cal R}_i$ and ${\cal T}_i$ defined as,
\bean
{\cal S}_i & = & \{ S_x^y \mid S_x^y \in {\cal S}, (p+1)i \leq y \leq (p+1)(i+1)-1 \}, \ \ i = 1,2,\ldots , q_1 \\
{\cal R}_i & = & \{ S_x^y \mid S_x^y \in {\cal S}_i, (p+1)(i+1) \leq x \leq d+1 \}, \ \ i = 1,2,\ldots , q_1 \\
{\cal T}_i & = & \{ S_x^y \mid S_x^y \in {\cal S}_i, (p+1)i+1 \leq x \leq (p+1)(i+1)-1 \}, \ \ i = 1,2,\ldots , q_1 
\eean
Note that ${\cal S}_i = {\cal R}_i \cup {\cal T}_i$. In similar lines of {\em Case 1(a)}, we proceed towards bounding the joint entropy $H({\cal S})$. We have,
\bea
H({\cal S}) &  \leq & \sum_{i=1}^{q_1} H({\cal R}_i) + \sum_{i=1}^{q_1} H({\cal T}_i) \\
& \leq & \sum_{i=1}^{q_1} \left(d-(i+1)(p+1)+2\right) \cdot  [2\beta - \theta + (p+1)\epsilon] + \sum_{i=1}^{q_1} \frac{(p+1)p\beta}{2} \label{eq:upper2a}
\eea
In the second inequality, we use \eqref{eq:rowsum2} of Prop.~\ref{prop:rowsum} to obtain the upper bound on $H({\cal R}_i)$. On the other hand, using Prop.~\ref{prop:colsum}, we also have,
\bea
H({\cal S}) &  \geq &  \sum_{i=p+1}^{(q_1+1)(p+1)-1} \min \{\alpha , (d-i+1)\beta \} - \epsilon \\
& = & \left[ \sum_{i=p+1}^{(q_1+1)(p+1)-1} (d-i+1)\beta \right] - \epsilon \label{eq:lower2a}
\eea
Matching the bounds in \eqref{eq:upper2a} and \eqref{eq:lower2a}, we must have
\bea
\epsilon & \geq & \frac{\left(d-\frac{(p+1)(q_1+3)}{2}+2 \right)q_1 \left[(p-1)\beta \ + \ \theta\right] }{\left(d-\frac{(p+1)(q_1+3)}{2}+2\right)q_1(p+1) \ + \ 1} \label{eq:eps2a}
\eea

{\em Case 2(b):} $q_1 =0$

\begin{figure}[h!]
\begin{center}
\includegraphics[width=2.8in]{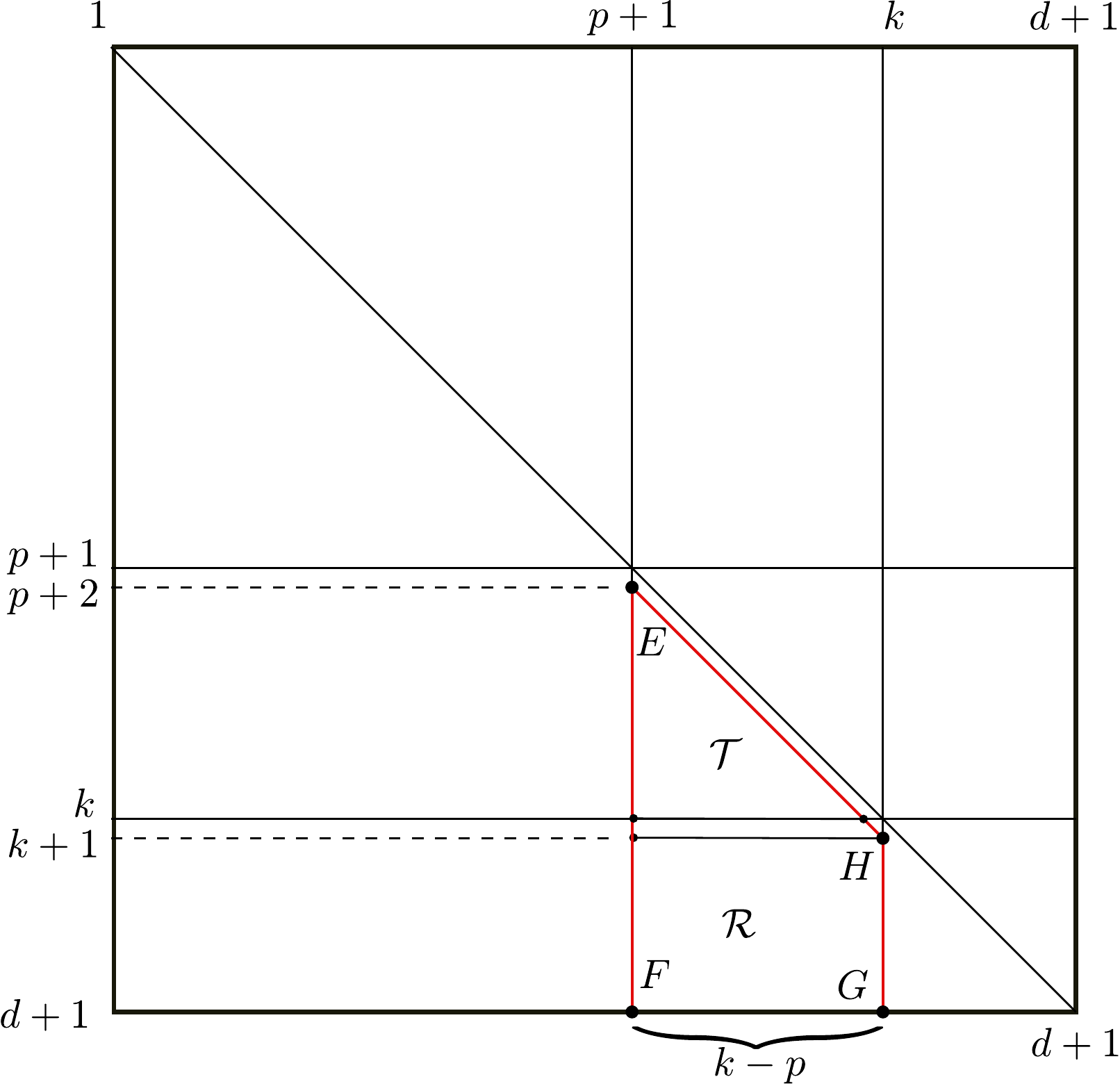}
\caption{The trapezium $EFGH$ marked in red in the repair matrix indicates the collection of random variables, ${\cal S}$, considered in {\em Case 2(b)}.}   \label{fig:case2b}
\end{center}
\end{figure}

Similar to the case of {\em Case 1(b)}, here we define
\bean
{\cal S} & = & \{ S_x^y \mid p+2 \leq x \leq d+1, p+1 \leq y \leq k , x > y \} \\
{\cal R} & = & \{ S_x^y \mid S_x^y \in {\cal S}, k+1 \leq x \leq d+1 \}, \\
{\cal T} & = & \{ S_x^y \mid S_x^y \in {\cal S}, p+2 \leq x \leq k \} 
\eean
Note that ${\cal S} = {\cal R} \cup {\cal T}$ . The pictorial representation of the above sets is given in Fig.~\ref{fig:case2b}, and ${\cal S}$ is represented by the trapezium $EFGH$. Using \eqref{eq:rowsum2} of Prop.~\ref{prop:rowsum}, we have
\bea
H({\cal S}) &  \leq & H({\cal R}) + H({\cal T}) \\
& \leq & (d-k+1) \cdot [2\beta - \theta + (k-p)\epsilon ] + \frac{(k-p-1)(k-p)\beta}{2} . \label{eq:upper2b}
\eea
On the other hand, using Prop.~\ref{prop:colsum},
\bea
H({\cal S}) &  \geq &  \sum_{i=p+1}^{k} \min \{\alpha , (d-i+1)\beta \} - \epsilon \\
& = & \left[ \sum_{i=p+1}^{k} (d-i+1)\beta \right] - \epsilon \label{eq:lower2b}
\eea
Matching the bounds in \eqref{eq:upper2b} and \eqref{eq:lower2b}, we must have
\bea
\epsilon & \geq & \frac{(d-k+1) \left[ (k-p-2)\beta + \theta \right]}{(d-k+1)(k-p) \ + \ 1} \label{eq:eps2b}
\eea

\epf

\end{example}


\end{document}